\documentclass[twoside,leqno,twocolumn]{article}

\usepackage[numbers,sort]{natbib}
\usepackage[letterpaper]{geometry}

\usepackage{latexsym,amsmath,textcomp,pifont,marvosym,wasysym,amssymb}
\usepackage{ltexpprt}
\usepackage{hyperref,cleveref}

\usepackage[algo2e,ruled,vlined,linesnumbered, noend]{algorithm2e}
\SetCommentSty{textit}
\DontPrintSemicolon
\SetAlCapHSkip{0pt}
\SetAlgoLined

\SetKwFor{ParallelFor}{for}{do in parallel}{}
\SetKwIF{If}{ElseIf}{Else}{if}{}{else if}{else}{end if}

\usepackage{tikz}
\usetikzlibrary{calc}

\usepackage{soul}
\usepackage{multirow}
\usepackage{bm}
\usepackage{booktabs}

\usepackage{pgfplots}
\usepgfplotslibrary{external}
\tikzsetexternalprefix{experiments/}

\usepackage{rotating}

\usepackage{mathtools}

\usepackage{multicol}

\newif\ifenablecomments
\enablecommentstrue
\newif\ifpdfplots
\pdfplotstrue

\newcommand{\mtkahypar}{Mt\nobreakdash-KaHyPar}
\newcommand{\mtkahip}{Mt\nobreakdash-KaHIP}
\newcommand{\mtmetis}{Mt\nobreakdash-Metis}

\newcommand{\kaminpar}{KaMinPar}

\newcommand{\jet}{Jet}

\newcommand{\Partition}{\ensuremath{\mathrm{\Pi}}}%
\newcommand{\cut}{\ensuremath{\mathrm{cut}}}%
\newcommand{\gain}[2]{\ensuremath{\operatorname{gain}(#2, #1)}}

\definecolor{fuchsiapink}{rgb}{1.0, 0.47, 1.0}

\definecolor{utahcrimson}{rgb}{0.83, 0.0, 0.25}

\definecolor{ao}{rgb}{0.0, 0.5, 0.0}

\newcommand{\plusplus}{\texttt{++}}

\newcommand{\gpp}[1]{g\plusplus#1}

\newcommand{\splitatcommas}[1]{%
  \begingroup
  \begingroup\lccode`~=`, \lowercase{\endgroup
    \edef~{\mathchar\the\mathcode`, \penalty0 \noexpand\hspace{0pt plus 1em}}%
  }\mathcode`,="8000 #1%
  \endgroup
}

\renewcommand{\epsilon}{\varepsilon}

\begin{document}

\newcommand\relatedversion{}

\title{\Large Parallel Unconstrained Local Search for Partitioning Irregular Graphs
       }

\author{
 Nikolai Maas\thanks{Karlsruhe Institute of Technology, Karlsruhe, Germany. \{nikolai.maas, lars.gottesbueren, daniel.seemaier\}@kit.edu} \and Lars Gottesbüren\footnotemark[1] \and Daniel Seemaier\footnotemark[1]
}

\date{}

\maketitle






\pagenumbering{arabic}

\begin{abstract} \small\baselineskip=9pt

We present new refinement heuristics for the balanced graph partitioning problem that break with an age-old rule.
Traditionally, local search only permits moves that keep the block sizes balanced (below a size constraint).
In this work, we demonstrate that admitting \emph{large} temporary balance violations drastically improves solution quality.
The effects are particularly strong on irregular instances such as social networks.
Designing efficient implementations of this general idea involves both careful selection of candidates for unconstrained moves as well as algorithms for rebalancing the solution later on.
We explore a wide array of design choices to achieve this, in addition to our third goal of high parallel scalability.
We present compelling experimental results, demonstrating that our parallel unconstrained local search techniques outperform the prior state of the art by a substantial margin.
Compared with four state-of-the-art solvers, our new technique finds 75\% of the best solutions on irregular graphs.
We achieve a 9.6\% improvement in edge cut over the next best competitor, while being only 7.7\% slower in the geometric mean.

\end{abstract}


\section{Introduction}

Balanced graph partitioning is a central problem in computer science with a huge array of applications.
The task is to divide the nodes $V$ of a graph $G=(V,E)$ into $k \in \mathbb{N}$ disjoint blocks $V_1, \dots, V_k \subseteq V$ of roughly equal size $|V_i| \leq (1+\varepsilon)\frac{|V|}{k} $ (the balance constraint) while minimizing the number of edges connecting different blocks $\sum_{i < j}^k |\{ \{ u, v \} \in E \mid u \in V_i, v \in V_j \}|$ (the edge cut).
A category of frequently studied applications is load-balanced data distribution while minimizing communication between parallel processors, which applies to a variety of scenarios such as distributed databases~\cite{schism}, graph processing, or scientific computing simulations.
These require extremely fast graph partitioning solvers with high solution quality.
Yet, balanced graph partitioning is NP-hard to approximate by a constant factor~\cite{APPROX-HARD}, which is why highly engineered heuristic solvers are used in practice.

The most important component in these solvers is local search to \emph{refine} a given partition.
In this setting the well-known Fiduccia-Mattheyses (FM) algorithm~\cite{FM} and its recently parallelized variants~\cite{MT-KAHIP, MT-KAHYPAR} are the most successful approaches.
Our contribution is a modification of parallel FM which achieves huge improvements on highly irregular instances such as social networks, web graphs, brain graphs and more.
Irregular graphs are characterized by highly skewed degree distributions, e.g., power-law distributions.

In each step, FM greedily moves a node to a different block, selecting the \emph{balance-preserving} move with \emph{highest gain} (reduction in edge cut).
Because negative gain moves are permitted to escape \emph{local minima}, it reverts back to the prefix with highest cumulative gain at the end of the search.
However, there is a second type of local minima: high gain moves that would violate the balance constraint.

In this work, we show how to escape the second type of local minima by using \emph{unconstrained} moves with some caution.
Our approach is based on estimating the \emph{cost} in terms of cut size to rebalance the partition later on.
This term is added as a penalty to the gain of balance-violating moves, thereby preventing moves which are unlikely to lead to an overall improvement.
Our experiments show that our new \emph{unconstrained FM} algorithm significantly outperforms prior refinement schemes on irregular graphs in terms of partition quality, while matching or slightly exceeding them on regular graphs.

In addition to refinement, state-of-the-art partitioners implement the multilevel scheme, which consists of the three phases \emph{coarsening}, \emph{initial partitioning} and \emph{uncoarsening/refinement}.
In the coarsening phase, it builds a hierarchy of increasingly smaller graphs by contracting node clusters, which aims to approximately preserve sparse cuts.
As such, the initial partition computed on the smallest graph is already a decent solution.
In the uncoarsening phase, the contractions are undone in reverse order, with local search refining the partition on each level.
Moving a node on coarse levels corresponds to moving a cluster of nodes on fine levels.
Thus, multilevel algorithms perform global optimization using local search.
Our contributions are in the refinement phase whereas we reuse prior work for the other two phases~\cite{MT-KAHYPAR-JOURNAL}.

\paragraph{Contributions}

We propose two parallel unconstrained local search algorithms: unconstrained FM and unconstrained label propagation, based on previous parallel constrained refinement algorithms~\cite{MT-KAHYPAR, MT-KAHIP}.
For unconstrained FM, we propose a new method to estimate the edge cut cost necessary to rebalance a balance-violating move later on and show how to maintain the estimation under concurrent node moves.
Moreover, we present an efficient parallel rebalancing algorithm that is designed to minimize the incurred cut increase.
Finally, we experimentally compare different ways to implement unconstrained refinement and evaluate our best performing algorithm against prior state-of-the-art solvers.

\paragraph{Results}

On irregular graphs we achieve a 9.6\% improvement in edge cut over the next best competitor, while being only 7.7\% slower in the geometric mean.
Out of a pool of five state-of-the-art-solvers, we find 75\% of the best solutions in total.
On regular graphs, our approach is also the best-performing solver overall.
While the margin is smaller, so is the running time overhead 
over the same baseline solver with constrained refinement.
As such, unconstrained refinement is suitable for both types of instances and thus constitutes the new state of the art for fast parallel refinement algorithms.

\section{Preliminaries}\label{s:preliminaries}

Let $G=(V,E,c,\omega)$ be an undirected graph with node weights $c:V \rightarrow \mathbb{N}_{>0}$ and edge
weights $\omega:E \rightarrow \mathbb{N}_{>0}$.
We extend $c$ and $\omega$ to sets in the natural way, i.e., $c(U) :=\sum_{v\in U} c(v)$ and $\omega(F) :=\sum_{e \in F} \omega(e)$.
Further, $\omega(v, U) \coloneqq \omega( \{ \{ v, u \} \in E \mid u \in U \} )$ is the summed edge weight between $v$ and $U$.
$N(v) \coloneqq \{ u \mid \{ v, u \} \in E \}$ denotes the neighbors of $v$ and
$I(v) \coloneqq \{ e \mid v \in e \}$ denotes the incident edges of $v$.
We say that $u \in N(v)$ is \emph{adjacent} to $v$.
The \emph{degree} of a node $v$ is $d(v) := |I(v)|$.

A \emph{$k$-way partition} of $G$ is a set of \emph{blocks} $\Partition \coloneqq \{V_1, \dots, V_k\}$ that partition $V$,
i.e., $V_1 \cup \cdots \cup V_k = V$ and $V_i \cap V_j = \emptyset$ for $i \ne j$.
We call $\Partition$ \emph{$\varepsilon$-balanced} if each block $V_i$ satisfies the \emph{balance constraint}:
$c(V_i) \leq L_{\max} := (1+\varepsilon) \lceil \frac{c(V)}{k} \rceil$ for some parameter $\mathrm{\varepsilon} > 0$.
Given parameters $\varepsilon$ and $k$, the \emph{graph partitioning problem} is to find an $\varepsilon$-balanced $k$-way partition $\Partition$ that minimizes $\cut(\Partition) := \sum_{i < j} \omega(E_{ij})$ (weight of all cut edges),
where $E_{ij} \coloneqq \{ \{ u, v \} \in E \mid u \in V_i, v \in V_j \}$.
We call a node $v \in V_i$ that has a neighbor $u \in V_j$ with $i \ne j$ a \emph{boundary node}.
We define $\Partition(v) \coloneqq V_i$ as the block containing $v$.
A move $m$ of $v$ removes $v$ from $\Partition(v)$ and assigns it to a different block,
thereby creating a new partition $\Partition'$.
Its \emph{gain} is defined as $\gain{\Partition}{m} \coloneqq \cut(\Partition) - \cut(\Partition')$.
We use $\Partition \circ m \coloneqq \Partition'$ to denote how moves operate on a partition.
Similarly, for a move sequence $\mathcal{M} = \left< m_1, \dots m_r \right>$, we use $\Partition \circ \mathcal{M} \coloneqq \Partition \circ m_1 \circ \dots \circ m_r$ for the created partition and $\gain{\Partition}{\mathcal{M}} \coloneqq \cut(\Partition) - \cut(\Partition \circ \mathcal{M})$ as the cumulative gain of $M$.

\section{Unconstrained Refinement}\label{s:unconstrained_refinement}

\begin{figure*}[t]
	\centering
	\includegraphics[width=0.85\textwidth]{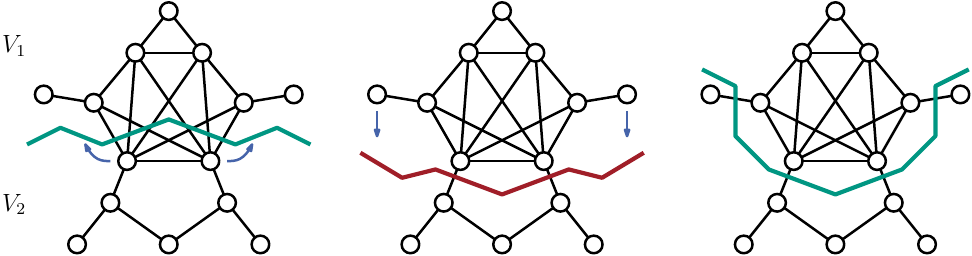}
	\caption{Illustration of unconstrained refinement.
		First, unconstrained moves are applied (left), which results in all high degree nodes being assigned to $V_1$ (center).
		Since the new partition is imbalanced, rebalancing is applied that moves low degree nodes to $V_2$.
		The resulting partition has a much smaller cut (right).}
	\label{fig:unconstrained_refinement}
\end{figure*}

In the following, we introduce the \emph{unconstrained refinement paradigm} as a framework for developing refinement techniques that work well on irregular graphs.
Unconstrained refinement can escape local minima that are hard to overcome for size-constrained refinement algorithms.
In this regard, previous techniques are limited by the following two aspects:
First, techniques based on label propagation or local search iteratively move single nodes while ensuring that the partition is balanced after every step~\cite{FM, MT-KAHYPAR, MT-METIS}.
Moving nodes to a block that already has the maximum weight
thus requires removing a similar number of nodes from the target block in advance.
Second, local search techniques typically consider only boundary nodes for moves~\cite{FM, MT-KAHIP},
thus excluding improvements that require moving block-internal nodes.
Even expensive techniques such as flow-based refinement restrict the explored solution space by using comparatively small subsets of nodes~\cite{KAHYPAR-HFC, MT-KAHYPAR-FLOWS}.

Unconstrained refinement overcomes these limitations by dividing each round of refinement into two phases.
In the first phase, \emph{unconstrained moves} are applied to the partition, i.e., moves that may violate the balance of the partition.
Even large imbalances are allowed in order to explore as much of the solution space as possible.
Afterwards, \emph{rebalancing} is applied to restore the balance of the partition.
Here, it is important that non-boundary nodes are also considered as move candidates.
Since rebalancing usually worsens the cut of the partition,
the results can be improved both with a high-quality rebalancing algorithm and by already considering the cut penalty when performing unconstrained moves.

Figure~\ref{fig:unconstrained_refinement} gives an example how unconstrained refinement can escape a local minimum.
Consider a highly irregular graph $G$ where the high degree nodes form a cluster $H$ and
each node in $H$ has many adjacent nodes with low degree.
Assume that the current partition $\{V_1, V_2\}$ divides $H$ into two subsets of similar size, including the adjacent low degree nodes in the according block
(this would be the natural way to partition $G$ into two connected blocks).
Unconstrained refinement moves all nodes in $H$ into the same block $V_1$ during the first phase.
Then, the balance is restored by moving low degree nodes from $V_1$ to $V_2$.
Due to the degree difference, this leads to a large improvement in cut size.

Achieving the same with moves that preserve the balance constraint requires a move sequence that swaps high and low degree nodes until all nodes in $H$ are moved to $V_1$.
However, such a sequence might not exist if the nodes have different weights.

Thus, unconstrained refinement allows to explore parts of the solution space which are hard to access otherwise.
However, it also comes with new challenges.
Generally, the goal is that the improvements of the first phase exceed the penalty incurred by the rebalancing.
But since this is hard to guarantee, a round of unconstrained refinement might actually worsen the cut of the partition.
To avoid this, algorithms for unconstrained refinement should include a \emph{rollback} mechanism such as restoring the best previously observed partition.
Furthermore, if the rebalancing reverts the previously applied moves, an oscillation might occur where the same moves are applied and undone for multiple rounds.
Possible strategies to avoid this include additional randomization of the performed moves or prohibiting certain moves for the following round.

\section{Related Work}\label{s:relatedWork}

There is a lot of literature on graph partitioning,
thus we refer the reader to surveys~\cite{GRAPH-OLD-SURVEY, GRAPH-RECENT-SURVEY, HYPERGRAPH-SURVEY, GRAPH-VERY-OLD-SURVEY} for a general overview.
The majority of modern hiqh-quality graph partitioners are based on the multilevel paradigm.
Publicly available shared-memory partitioners include \mtkahip~\cite{MT-KAHIP}, \mtmetis~\cite{MT-METIS} and \kaminpar~\cite{KAMINPAR}.
Further, \mtkahypar~\cite{MT-KAHYPAR, MT-KAHYPAR-Q, MT-KAHYPAR-DET} is a shared-memory hypergraph partitioner that includes a graph configuration with improved running time~\cite{MT-KAHYPAR-JOURNAL}.

\paragraph{Refinement and Parallelization}

For the refinement phase, multilevel partitioners use a variety of local search techniques~\cite{MT-KAHIP, MT-METIS, MT-KAHYPAR}.
\emph{Size-constrained label propagation} visits all nodes in parallel and greedily moves each node to the block with highest positive gain~\cite{LABEL_PROPAGATION}.
The \emph{greedy refinement} used by \mtmetis~\cite{MT-METIS} extends label propagation with thread-local priority queues,
always selecting the highest positive gain move.
Similarly, the \emph{Fiduccia-Mattheyses (FM)} algorithm repeatedly performs the move with highest gain but also allows negative gain moves~\cite{FM}.
Then, the best observed solution is applied to the partition.
Moves that violate the balance constraint are prohibited in FM,
but with negative gain moves it still has the potential to escape local minima.
However, the FM heuristic is difficult to parallelize since it requires a serial move order.
While some partitioners perform 2-way FM refinement on independent block pairs~\cite{SCOTCH, KAPPA}, this provides only limited parallelism.
Instead, the parallel $k$-way FM variant used in \mtkahip{} and \mtkahypar{} relaxes the requirement to perform moves in serial order~\cite{MT-KAHIP, MT-KAHYPAR}.
The algorithm uses parallel localized searches and afterwards combines the results into a global move sequence.

\paragraph{Prior Work on Unconstrained Refinement}

Previous work on single-level hypergraph partitioning 
already observed that temporary balance violations can improve the partition quality.
Caldwell et al. propose to use multiple refinement passes for inputs with non-uniform node weights~\cite{VARYING-NODE-WEIGHTS},
where the first pass uses a relaxed balance constraint that ensures every node is movable.
Each following pass tightens the balance constraint and applies greedy rebalancing before the refinement.

Dutt and Theny consider a more complicated approach~\cite{INTERMEDIATE-RELAXATION},
where temporary balance violations are allowed if an \emph{estimator} predicts that the overall gain is positive after rebalancing.
They cite being able to move nodes with very high weight as their competitive advantage.
In addition, the authors mention that constrained algorithms might lock a cluster in the cut,
while a temporary balance violation allows to move the complete cluster.
Their estimator calculates a rating by finding a move set that restores balance.
Additionally, it includes a look-ahead mechanism which considers edge cut reductions through additional balance-violating moves in the future.
This allows to draw clusters of nodes across the cut, but is also expensive to calculate.
Therefore, they propose a heuristic variant using multiple factors,
most importantly the gain difference in relation to the highest gain move of the target block.
Their approach is designed for a single-level algorithm on regular instances and hard to parallelize (due to the global priority queue and unclear ownership of nodes).
Our work follows similar ideas in that we penalize balance-violating moves.
Yet, we use a simpler and more efficient penalty based on the rebalancing cost for the next move,
which is also compatible with prior parallelization techniques~\cite{MT-KAHIP, MT-KAHYPAR}.

Furthermore, cluster move techniques lost their importance with the rising dominance of the multilevel paradigm.
Multilevel partitioners went back to simple refinement algorithms,
citing the ability to move whole clusters of nodes at coarser levels as the main device to overcome local minima~\cite{JOSTLE, METIS}.
Some multilevel systems allow small amounts of imbalance on coarser levels, and increasingly tighten the constraint during uncoarsening until the desired balance is reached on the finest level~\cite{JOSTLE-RELAXED-BALANCE, KAHIP-RELAXED-BALANCE}.
It was observed that this can improve result quality in combination with FM refinement~\cite{KAHIP-RELAXED-BALANCE}.
However, the effect is restricted to small imbalances on coarse levels and thus no replacement for unconstrained refinement.
To the best of our knowledge,
there is only one recent work that allows large temporary balance violations as part of the refinement algorithm~\cite{JET}.
The \emph{\jet{}} algorithm precomputes gains for all boundary nodes and applies promising moves in parallel.
Afterwards, a separate rebalancing algorithm is applied.
Here, allowing balance violations enables a degree of parallelism that is suitable for the GPU.
In addition, the authors observed that \jet{} achieves better cut sizes than size-constrained label propagation and even outperforms FM refinement in some cases~\cite{JET}.

\section{Unconstrained $k$-way FM Local Search}
\label{s:ufm}

In the following, we present an unconstrained parallel version of the FM local search~\cite{FM}.
We apply an \emph{approximate penalty} to unconstrained moves which represents the cost of rebalancing the partition afterwards.
The actual rebalancing is performed in a separate step.
Then, unconstrained moves and rebalancing moves are combined into a new move sequence
and the best prefix is applied to the partition.

We recap the sequential FM algorithm in Section~\ref{s:classical_fm} and its shared-memory parallelization~\cite{MT-KAHIP, MT-KAHYPAR} in Section~\ref{s:parallelization}.
The remaining sections describe the details of our new unconstrained FM algorithm.
We use the high-level pseudocode in Algorithm~\ref{algo:localized-fm-sketch} to follow along the description, and highlight in blue where we adapted the algorithm.

\subsection{Sequential FM}
\label{s:classical_fm}

The FM algorithm works in two phases:
First, it constructs a sequence of moves $\langle m_1, \dots, m_l \rangle$ by repeatedly applying a \emph{balance-preserving} move with highest gain.
Afterwards, it reverts to the prefix with highest cumulative gain in the move sequence.
Let $\Partition^i$ be the partition after applying $\langle m_1, \dots, m_{i-1} \rangle$.
The prefix is chosen such that $i$ maximizes $\sum_{j=1}^i \gain{\Partition^{j}}{m_j}$.
Since negative gain moves are allowed during the construction, FM can escape some local optima~\cite{FM}.
However,
a different type of local optimum occurs if the moves with highest gain violate the balance constraint.
By allowing \emph{unconstrained} moves during the first phase,
our FM variant can overcome both kinds of local optima.

\definecolor{kitblue}{RGB}{70, 100, 170}

\begin{algorithm2e}[t]
	\caption{\textcolor{kitblue}{Unconstrained} Parallel FM}\label{algo:localized-fm-sketch}
	\SetKwFor{ParallelDo}{parallel do}{}{}
	\textcolor{kitblue}{$V_r \gets \{ v \in V \mid \omega(v, \Partition(v)) \ge t \cdot \omega(I(v)) \}$} \;
	\textcolor{kitblue}{$\FuncSty{initializePenaltyEstimation}(V_r)$} \;
	$\mathcal{M} \gets \langle \rangle$ \tcp*[r]{global move sequence}
	\ParallelDo(){}{
		$\mathcal{M}_l \gets \langle \rangle$ \tcp*[r]{local move sequence}
		initialize with random boundary nodes \;
		\While() {not done } {
			perform local move $m$ that maximizes $\gain{\Partition}{m}$ \textcolor{kitblue}{$- \operatorname{penalty}(m, \Partition)$}
			\;
			update and acquire neighbors \;
			$\mathcal{M}_l \gets \FuncSty{append}(\mathcal{M}_l, m)$\;
			\textcolor{kitblue}{
			\If() {moved node in $V_r$}{
				adjust penalty estimation for origin block of $m$\;
			}
			}
			\If () {$\gain{\Partition}{\mathcal{M}_l}$ \textcolor{kitblue}{$- \operatorname{penalty}(\mathcal{M}_l, \Partition)$} $> 0$} {
				$\mathcal{M} \gets \FuncSty{append}(\mathcal{M}, \mathcal{M}_l)$\;
				$\Partition \gets \Partition \circ \mathcal{M}_l$ \;
				$\mathcal{M}_l \gets \langle \rangle$ \;
			}
		}
	}
	\textcolor{kitblue}{
	\If() {$\Partition$ is imbalanced}{
		$\mathcal{R} \gets \FuncSty{rebalance}(\ArgSty{\Partition})$\;
		$\mathcal{M} \gets$ interleave $\mathcal{M}$ and $\mathcal{R}$ \;
	}
	}
	recalculate gains in $\mathcal{M}$ and apply best prefix \;
\end{algorithm2e}

\subsection{Shared-memory Parallel FM}
\label{s:parallelization}

Parallel $k$-way FM uses non-overlapping \emph{localized} searches on multiple threads for the first phase of the FM algorithm~\cite{MT-KAHIP, MT-KAHYPAR}.
Each search maintains a local priority queue that is initialized with a small number of boundary nodes.
The searches expand by acquiring ownership of nodes that are adjacent to performed moves,
provided that no other thread already owns the node (see line 9).
Each search maintains a thread-local partition and move sequence,
updating the shared partition only if the local move sequence provides a net improvement (see line 13-16).
In the second phase, the algorithm constructs a global move sequence and recalculates the gains of all moves according to their position in the sequence.
Then, the partition is reverted to the best global prefix.
We refer to Ref.~\cite{MT-KAHYPAR-JOURNAL} for more details on the engineering aspects of this algorithm.

\subsection{Calculating Approximate Penalties}
\label{s:penalties}

For each move that violates the balance constraint, we add an approximate penalty to the gain of the move (see line~8).
This affects both its position in the local priority queue and the computation of the best local prefix.

The penalty is based on the gains and weights of nodes that might be used to rebalance after the balance-violating move.
We group nodes by their current block and their weight ratio (see below) into exponentially spaced buckets.
Thereby, we can access more promising rebalancing moves first, but avoid the overhead of sorting.
We say that a node $v$ is \emph{available for rebalancing} if $\omega(v, \Partition(v)) \ge t \cdot \omega(I(v))$, where $t$ is a tuning parameter (experimentally chosen as $t = 0.7$).
This means that at most a fraction of $1-t$ of the incident edges of $v$ is connected to the boundary of the block.
At the start of each FM round we initialize the buckets using the nodes that are available for rebalancing (see line 1 and 2).
The bucket of $v$ is determined by the current block of $v$ and a slot that is given by $\mathrm{slot}(v) \coloneqq \lceil  \log_{\frac{3}{2}} \frac{\omega(v, \Partition(v))}{c(v)} \rceil$.
Thus, the slot of a node represents the ratio of its internal edge weight to node weight.
Note that we use base $\frac{3}{2}$ for the bucket width to achieve better accuracy than with base $2$.
Now, assume that $m$ moves a node $u$ to an overloaded block $V_j$ (i.e., $c(V_j \cup \{u\}) > L_{max}$).
To simulate a rebalancing of $m$ we consider an according proportion (with regards to the node weight) of the bucket slot that matches the imbalance of $V_j \cup \{u\}$.
More formally, let $B_{i,j}$ be the nodes assigned to the bucket with slot $i$ and block $V_j$.
We choose $l$ as the minimal slot such that $\sum_{i=0}^l c(B_{i,j}) \ge c(V_j \cup \{u\} ) - L_{max}$ holds\nolinebreak
\footnote{If no such $l$ exists we forbid the move.}.
Then, based on the weight ratio $\frac{3}{2}^l$ of the slot,
the penalty for node $u$ is $\mathrm{pen}(m, \Partition) \coloneqq \mathrm{pen}(u, V_j) \coloneqq \frac{3}{2}^l c(u)$.
This is an estimate for the increase in cut size when nodes from slot $l$ are used to compensate the additional weight of $u$ in $V_j$.


\paragraph{Performance Guarantees}

As we show in the following lemma,
this penalty provides a bound on the rebalancing cost in a simplified scenario with unit node weights and where no node leaves an overloaded block.
While the lemma is formulated for one overloaded block, the argument also works for multiple overloaded blocks as their rebalancing nodes are disjoint.

\begin{Lemma}
	\label{lemma:penalty}

	Consider a graph $G$ with unit node weights, a balanced partition $\{V_1, \dots, V_k\}$ of $G$ and a set of nodes $U = \{u_1, \dots, u_j\}$ that are moved to $V_1$.
	Let $U^i \coloneqq \{u_1, \dots, u_i\}$ for $i \le j$.
	Then, the partition $\Partition \coloneqq \{V_1 \cup U, V_2 \setminus U, \dots, V_k \setminus U\}$ can be rebalanced while increasing the cut by at most $\frac{1}{t} \sum_{i=1}^j \mathrm{pen}(u_i, V_1 \cup U^{i-1})$.
\end{Lemma}

\begin{proof}
	We can rebalance $\Partition$ by moving $j$ nodes from $V_1$ to another block.
	Choose $j$ nodes $b_1, \dots, b_j$ from $B_{1, 1} \cup B_{2, 1} \cup \cdots$, using the smallest available slot for each node.
	Moving $b_1, \dots, b_j$  increases the cut at most by their incident edge weight.
	Therefore, the rebalancing cost is bounded by
	$ \sum_{i=1}^j \omega(I(b_i)) \le \frac{1}{t} \sum_{i=1}^j \omega(b_i, V_1) \le \frac{1}{t} \sum_{i=1}^j \frac{3}{2}^{\mathrm{slot}(b_i)} = \frac{1}{t} \sum_{i=1}^j \mathrm{pen}(u_i, V_1 \cup U^{i-1}) $.
	The last equality holds since the exponent used for the penalty is always the index of the smallest available slot.
\end{proof}

Therefore, we can derive an upper bound for the rebalancing cost
while deferring the actual rebalancing to a later step.
Note that the bound could be strengthened by removing the factor $\frac{1}{t}$ if we use $\omega(I(v))$ instead of $\omega(v, \Partition(v))$ for determining the slot of $v$.
However, the penalty is already overly pessimistic in practice, thus it is not desirable to make it even more pessimistic for boundary nodes.

\paragraph{Parallel Updates}

Determining penalties in a parallel setting incurs additional difficulties.
A node $v$ that was marked as available for rebalancing in the beginning might be moved during the FM round.
Thus, it can no longer be used to rebalance and consequently should be eliminated from the penalty calculation.
Correctly updating the buckets is non-trivial
as the move is (temporarily) only visible to the localized search that owns the node and might be reverted later.
Instead, we use a simpler approach where each node remains in its initial bucket.
We achieve a similar effect to updating the buckets
by tracking a \emph{virtual weight delta} $\tilde{c}_j$ for each block $c(V_j)$
and use $c(V_j) + \tilde{c}_j$ instead of $c(V_j)$ to determine the bucket slot in the penalty calculation.
$\tilde{c}_j$ is initially zero and increases by $c(v)$ if a node $v$ that is available for rebalancing is removed from $V_j$.
In Algorithm~\ref{algo:localized-fm-sketch} this happens in line 12.
This compensates that $v$ is no longer available and thus avoids too small penalties.

\subsection{Rebalancing and Global Move Sequence}
\label{s:move_sequence}

After finding moves in the localized search using our approximate penalty,
the next step is to combine them into a global move sequence.
When a thread-local search finds a new best prefix $\langle m_1, \dots, m_i \rangle$ of its local move sequence
that maximizes $\sum_{j=1}^i \gain{\Partition^{j-1}}{m_j} - \mathrm{pen}(m_j, \Partition^{j-1})$,
the prefix is immediately applied to the shared partition (see line 14 and 15 in Algorithm~\ref{algo:localized-fm-sketch}).
We linearize the local move sequences into a global move order $\mathcal{M}$, as described in Ref.~\cite{MT-KAHYPAR}.
If the partition resulting from $\mathcal{M}$ is imbalanced, we apply our rebalancing algorithm (see line 18).
We record all rebalancing moves
and group them by the original block of the moved node into the sequences $\mathcal{R}_1, \dots, \mathcal{R}_k$,
using the order in which they were performed.
Since it is possible that almost every prefix of $\mathcal{M}$ is imbalanced,
we now construct an \emph{interleaved} move sequence $\mathcal{I}$.
Let $\mathcal{I}^j$ be the currently constructed prefix of $\mathcal{I} \coloneqq \mathcal{I}^{|\mathcal{M}|}$ with $\mathcal{I}^0 \coloneqq \langle \rangle$,
and let $\Partition^j \coloneqq \Partition \circ \mathcal{I}^j$ be the according partition.
In each step, we append the next move  $m^{j+1} \in \mathcal{M} \setminus \mathcal{I}^j$.
If $\Partition^j \circ m^{j+1}$ is imbalanced,
we use the rebalancing moves $\mathcal{R}_i$ of the currently overloaded block to immediately restore balance.
Formally, we choose a minimal prefix $\langle r_1^{j+1}, \dots, r_l^{j+1} \rangle$ of $\mathcal{R}_i \setminus \mathcal{I}^j$
such that $\Partition^j \circ m^{j+1} \circ r_1^{j+1} \circ \cdots \circ r_l^{j+1}$ is balanced.
$\mathcal{I}^{j+1}$ is the concatenation of $\mathcal{I}^{j}$ and $\langle m^{j+1}, r_1^{j+1}, \dots, r_l^{j+1} \rangle$.

Since the originally computed gains are incorrect for $\mathcal{I}$ (due to the changed move order),
we now calculate exact gains using a parallel gain recalculation algorithm~\cite{MT-KAHYPAR}\nolinebreak
\footnote{
	The parallel gain recalculation has the precondition that each node is moved at most once,
	which might be violated by the rebalancing moves.
	Therefore, for any node that is moved twice we first combine the two moves into a single move.
}.
Afterwards, we revert to the prefix of $\mathcal{I}$ with highest cumulative gain.

\subsection{Integration into the Overall Algorithm}
\label{s:integration}

We use a total of ten rounds of FM local search.
However, the unconstrained FM algorithm comes with trade-offs that need to be considered when designing the overall algorithm.
First, it is often slower than constrained FM since it performs more moves and subsequent rebalancing is required.
Second, whether unconstrained FM achieves significantly improved quality depends on the structure of the graph.
In addition to this, the approximate penalty for unconstrained moves tends to be too pessimistic.

Our first technique for addressing this consists of gradually changing the severity of the calculated penalties.
In the $i$'th round of unconstrained FM local search we multiply all penalties with a factor $\tau_i \le 1$.
We start with a small value for $\tau_1$
and increase it over the following rounds using linear interpolation between $\tau_1$ and $\tau_{final} = 1$.
Also, we apply at least one round of the constrained FM algorithm after the unconstrained rounds.
The effect of this is that the early rounds
perform more unconstrained moves and are thus more likely to find drastic changes to the partition.
Later rounds perform unconstrained moves only if they are clearly beneficial,
thereby converging to a specific solution (similar to, e.g., the simulated annealing metaheuristic).
Additionally, we switch from unconstrained FM to constrained FM if the improvement of one round relative to the overall cut is below a certain threshold (experimentally determined as $0.002$).
This avoids the running time overhead of unconstrained FM in cases where it does not provide improved quality.

To further minimize the overhead, 
we implemented a technique for deciding dynamically whether the unconstrained or the constrained FM algorithm is used.
We use the first two rounds to decide between the two by running one round of constrained FM and afterwards one round of unconstrained FM with $\tau_0 = \frac{1}{2}$.
If the improvement found by the second round is higher than that of the first round,
we use unconstrained FM as described above.
Otherwise, we use the constrained FM algorithm.

\section{Unconstrained Label Propagation}

Label propagation is a fast refinement technique
which is widely used in parallel partitioning algorithms~\cite{MT-KAHIP, MT-KAHYPAR, KAMINPAR}.
Since only positive gain moves are performed, \emph{size-constrained} label propagation can not escape from local optima.
However, \emph{unconstrained} label propagation might at least escape from one kind of local optimum via a temporary balance violation.
In practice, we observed that it actually provides a quality improvement in combination with unconstrained FM.

Our algorithm uses multiple rounds and maintains a set of \emph{active} nodes $V_a$.
Initially, $V_a$ contains all boundary nodes.
In each round, we iterate in parallel over $u \in V_a$ and move $u$ to the neighboring block with the highest positive gain, if any.
In contrast to size-constrained label propagation, we explicitly allow moves that violate the balance constraint.
We then apply our rebalancing algorithm if the partition is imbalanced.
Since this might increase the cut size, we check whether the round resulted in a net improvement.
If not, we apply a rollback that restores the partition state at the beginning of the round and terminate.
Otherwise, we determine the new set $V_a$ of active nodes as follows.
For each moved node,
we add its neighborhood to $V_a$.
However, we only add nodes that were not moved themselves, in order to avoid oscillations.
We stop after five rounds, if $V_a$ is empty, or if the last round's net improvement is below a threshold (we use $0.001$) of the total cut size.
Based on our observations, this leads to fast convergence in practice.

The extensive pruning of active nodes and early termination are a deliberate trade-off that favors speed over quality.
This is less desirable if unconstrained label propagation is the only refinement algorithm
(compare Ref.~\cite{JET}).
However, the algorithm is designed to be used before our unconstrained FM algorithm.
Since the latter is capable of finding non-trivial improvements, it is unnecessary to design the label propagation algorithm for high quality.

\section{Rebalancing}
\label{s:rebalancing}

In the following, we present our parallel rebalancing algorithm, which takes an imbalanced partition and balances it by moving nodes out of the overloaded blocks.
We opted for a parallel priority-based approach, where all threads poll moves from a central concurrent priority queue.
This is motivated by the following observation.
Almost all balance-improving moves have negative gain.
Therefore, it is important to move the most promising nodes first and only those, terminating as soon as balance is achieved.
We use multi-queues~\cite{multiqueues-spaa, multiqueues-esa}, a concurrent relaxed priority queue with good practical performance characteristics that is simple to implement.

\subsection{Overview}

The rebalancing algorithm consists of two phases (insertion and moving).
In the insertion phase, we insert \emph{all} nodes in overloaded blocks into the concurrent priority queue (PQ) as potential candidates to be removed from the overloaded blocks.
This includes nodes that were just moved, in case it is more beneficial to move some of them back.
In the moving phase, we extract nodes from the PQ, move them to their preferred target block, and update their neighbors' priorities to reflect the move.
During rebalancing, only moves into non-overloaded blocks are permitted.
Nodes that are not connected to a valid target block are moved to the block with lowest total node weight.

For a node $u$, we consider the move $m$ with best gain that preserves the balance of the target block.
We use $\gain{\Partition}{m}/c(u)$ as the priority of $u$, if $\gain{\Partition}{m} < 0$ (this is the expected case) and $\gain{\Partition}{m} \cdot c(u)$ if $\gain{\Partition}{m} \geq 0$.
This function combines the two goals of progressing quickly to a balanced state and not worsening the solution more than necessary.
We use the concurrent gain table that is already used by parallel localized FM~\cite{MT-KAHYPAR} to look up $\gain{\Partition}{m}$ and update the move priorities.

Once a block is no longer overloaded, we skip its nodes when they are extracted from the PQ.
Moreover, we stop once no overloaded blocks are left.
We then serialize the moves in the order in which they were performed and group them by their origin block.

We use atomic fetch-and-add instructions to update the block weights, detect when they are no longer overloaded, and prevent new blocks from accidentally becoming overloaded.
While this exhibits high contention on the block weights, the main bottlenecks are gain updates and priority queue updates.

\subsection{Implementation Details}

In this section, we present implementation details of the rebalancing algorithm.
We begin with a description of multi-queues~\cite{multiqueues-esa} and then expand on the specialized implementation for our scenario.

\paragraph{Relaxed Priority Queue}

Let $\tau$ be the number of threads and let $C$ be a small constant ($C=2$ is a recommended value).
Multi-queues~\cite{multiqueues-spaa, multiqueues-esa} use $C  \cdot \tau$ sequential priority queues (SPQ), which are guarded with locks.
To retrieve the \emph{relaxed} maximum element, we compare the maximum elements from two randomly selected SPQs.
For insertion, we send the element to a random SPQ.
With all operations we only use \emph{try-lock}, never insisting on taking a lock.
If the try-lock fails, we retry the operation with different randomization.
Hence, all operations are wait-free.
Since we have $C$ times as many PQs as threads, each try-lock succeeds with probability at least $1- \frac{1}{C}$.

\paragraph{Neighbor Updates}

After we move a node, we inspect its neighbors to update their gain in the gain table and adjust their priority in the multi-queue.
We lock nodes to prevent that multiple threads update the same node's priority concurrently.
Additionally, we lock a node before it is extracted from the PQ, to avoid updating removed elements.
If we successfully take a neighbor's lock, we compute and save its preferred target block and corresponding gain, but do not update the priority just yet (see next paragraph).
If we do not get the lock, we know that a different thread will perform the operation.
Since it uses the gain from the globally shared gain table, the updated value will be (roughly) the same.

To reduce contention on the SPQ locks, we group all neighbors of the last moved node by the SPQ they are in.
We cycle through the not yet updated SPQs and try to lock them.
Once locked, we perform a bulk update of the locked nodes in this SPQ, and only then release the node locks.
If an SPQ try-lock fails, we simply move on to the next SPQ, until all updates have been performed.

\paragraph{Finding the Next Move}

When finding the next move, we perform a \FuncSty{tryDeleteMax()} operation on the PQ.
We use a technique known from parallel FM to avoid inaccurate gains caused by race conditions.
After selecting a node from the PQ, we check its gain against the gain table.
If that changed for the worse, we update the node's priority and retrieve a new relaxed maximum priority node.
We fuse the \FuncSty{deleteMax()} operation and the node locking with this double-check for the following reason.
Our scenario permits a simplified \FuncSty{empty()} check because insertions and deletions do not happen concurrently.
While a straight-forward implementation of the gain double-check may first delete and then reinsert an element, this conflicts with the simplified \FuncSty{empty()} check.
See also~\cite{multiqueues-esa} for a discussion of the intricacies of \FuncSty{empty()} checks when insertions and deletions are concurrent.

After selecting the appropriate SPQ from two randomly chosen ones and acquiring its lock, we try to acquire the node's lock (for the move) and run the double-check.
If both steps succeed we perform the \FuncSty{deleteMax()}, release the SPQ lock and then return the node.
If the node-lock fails, we unlock the SPQ and try again by drawing two new SPQs.
If the double-check fails, we adjust the node's priority, release the SPQ lock and try again by drawing two new SPQs.

\section{Experiments}\label{s:experiments}

We start our experimental evaluation by making a foray into the design space of unconstrained refinement in~\Cref{subsec:design_space}.
In~\Cref{subsec:scaling}, we then present strong scalability results for the rebalancing algorithm.
Finally, in \Cref{subsec:solution_quality,subsec:time}, we compare our new refinement algorithms with the state-of-the-art parallel multilevel partitioners \mtmetis~\cite{MT-METIS-REFINEMENT} (with hill-scanning and two-hop coarsening), \kaminpar~\cite{KAMINPAR} (fast configuration) and \mtkahypar~\cite{MT-KAHYPAR} (default configuration: $\log(n)$-level coarsening and no flow-based refinement), as well as \jet~\cite{JET}.
We excluded \mtkahip~\cite{MT-KAHIP} since it is stricly outperformed by \mtkahypar{}~\cite{MT-KAHYPAR-JOURNAL}.
As the source code for Jet is not publicly available, we implemented its refinement in \mtkahypar{} with guidance from the authors of Jet. An added benefit of this approach is that we compare the refinement without differences due to surrounding multilevel components, namely coarsening and initial partitioning.

\subsection{Experimental Setup}

We implemented both our new refinement algorithms and \jet{} in the \mtkahypar{} framework, as the Jet source code is unavailable.
This has the additional benefit that we avoid measuring differences from components other than refinement.
Our source code is available at \url{https://github.com/kahypar/mt-kahypar/tree/unconstrained-refinement} and \url{https://github.com/kahypar/mt-kahypar/tree/jet-refiner}.

All experiments are run on an AMD EPYC Rome 7702P (one socket with $64$ cores) running at $2.0$--$3.35$ GHz with $1024$GB RAM and 256MB L3 cache.
The code is parallelized using TBB~\cite{TBB} and compiled using \gpp{ 9.4} with full and native architecture optimizations.
All runs are performed on 64 cores.

\paragraph{Benchmark Sets}

We compiled two benchmark sets I and R\footnote{
    The benchmark sets and experimental results are available at \url{https://zenodo.org/records/15386627}
}, consisting of large graphs from the SuiteSparse Matrix Collection~\cite{SPM} and Network Repository~\cite{NetworkRepository},
which are categorized as either highly \textbf{i}rregular (such as social networks) or fairly \textbf{r}egular (such as meshes originating from scientific simulations).
This split was made to show that unconstrained local search yields enormous quality benefits on set I, whereas it performs similarly to prior work on set R.
All graphs and their categories are listed in \Cref{tab:graphs}.
Set I contains 38 graphs ranging from 5.4 million edges to 1.8 billion edges.
It includes social networks, web graphs, wiki graphs and graphs that model the human brain.
We also added a set of graphs created by compressing texts~\cite{TEXT-RECOMPRESSION} from the Pizza\&Chili corpus~\cite{PIZZA-CHILI-TEXTS}, as well as artificial instances from graph models with skewed degree distribution.
Set R contains 33 graphs ranging from 12.7 million edges to 575 million edges.
It includes road graphs, biological graphs modeling amino acids, graphs originating from non-linear optimization problems, mesh graphs from finite element models, semiconductor circuits and artificial graphs from models with regular degree distribution.
All graphs are unweighted except for the text compression graphs, which have edge weights.

The split into regular and irregular was determined by statistics on the node degrees, specifically the standard deviation divided by the mean degree.
For set R this ratio ranges from 0.026 (afshell10) to 1.37 (stokes), whereas for set I it ranges from 1.21 (bn-M87117515) to 1354 (mavi).
We decided to keep graphs from the same category together in the same benchmark set, which explains why the ranges overlap for the categories semi-conductor and brain graphs.

\begin{table*}[t]
	\caption{Graphs in set I (upper half) and set R (lower half), grouped into classes based on their origin. Within each class, the graphs are ordered by increasing number of edges.}
	\label{tab:graphs}

	\vspace{6pt}
	\begin{tabular}{lrp{0.775\textwidth}}
		Class & \# & Graphs\\
		\midrule
		Social & 8 & imdb2021, flickr-und, livejournal, hollywood, orkut, sinaweibo, twitter2010, friendster \\
		Web & 7 & mavi201512020000, indochina2004, arabic2005, uk2005, webbase2001, it2004, sk2005 \\
		Wiki & 5 & eswiki2013, itwiki2013, frwiki2013, dewiki2013, enwiki2022 \\
		Brain & 5 & bn-M87117515, bn-M87123142, bn-M87122310, bn-M87126525, bn-M87128519-1 \\
		Compression & 6 & sources1GB-7, sources1GB-9, english1GB-7, dna1GB-9, proteins1GB-7, proteins1GB-9 \\
		Artificial & 7 & rmat-n16m24, kron-g500n19, rhg-n23d4, kron-g500n20, kron-g500n21, rhg-n23d20, rmat-n25m28 \\
		\midrule
		Road & 2 & asia-osm, europe-osm \\
		Biology & 6 & cage15, kmer-V2a, kmer-U1a, kmer-P1a, kmer-A2a, kmer-V1r \\
		Optimization & 2 & nlpkkt200, nlpkkt240 \\
		Finite element & 16 & ldoor, afshell10, boneS10, Hook1498, Geo1438, Serena, audikw, channelb050, LongCoup-dt6, dielFilterV3, MLGeer, Flan1565, Bump2911, CubeCoup-dt6, HV15R, Queen4147 \\
		Semiconductor & 4 & nv2, vas-stokes2M, vas-stokes4M, stokes \\
		Artificial & 3 & delaunay-n24, rgg-n263d, rgg-n26
	\end{tabular}
\end{table*}

\paragraph{Methodology}

We use an imbalance of $\varepsilon = 0.03$, which is a standard parameter in the literature, $k \in \{2, 4, 8, 11, 16, 17, 23, 32\}$, and ten random seeds for each instance (combination of graph and $k$).
For each instance, we aggregate running times and cut size using the arithmetic mean over all seeds.
To further aggregate over multiple instances, we use the geometric mean for absolute running times and the quality relative to the best solution.
Runs with imbalanced partitions are not excluded from aggregated running times.
For runs that exceeded the time limit, we use the time limit itself in the aggregates.
In plots, we mark instances where \emph{all} runs of that algorithm timed out our could not produce a balanced partition with \ding{55}.

\paragraph{Performance Profiles}

To compare the cut sizes of different algorithms, we use \emph{performance profiles}~\cite{PERFORMANCE-PROFILES}.
Let $\mathcal{A}$ be the set of algorithms we want to compare, $\mathcal{I}$ the set of instances, and $q_{A}(I)$ the cut of algorithm $A \in \mathcal{A}$ on instance $I \in \mathcal{I}$.
For each algorithm $A$, we plot the fraction of instances ($y$-axis) for which $q_A(I) \leq \tau \cdot \min_{A' \in \mathcal{A}}q_{A'}(I)$, where $\tau$ is on the $x$-axis.
Achieving higher fractions at lower $\tau$-values is considered better.
For $\tau = 1$, the $y$-value indicates the percentage of instances for which an algorithm performs best.

\subsection{Design Space}\label{subsec:design_space}

\begin{figure*}
	\begin{minipage}{0.49\textwidth}
	\ifpdfplots
		\includegraphics{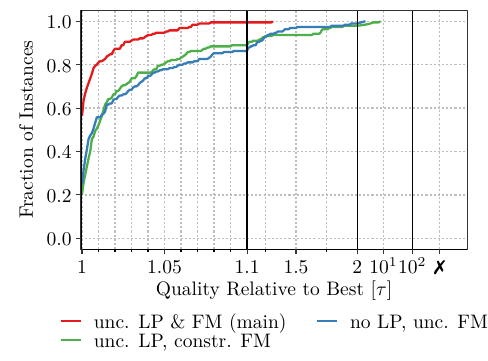}
	\else
		\tikzsetnextfilename{irregular_algo_combinations}%
		\input{tikz/irregular_algo_combinations}%
	\fi

	\end{minipage}
	\hfill
	\begin{minipage}{0.49\textwidth}
	\ifpdfplots
		\includegraphics{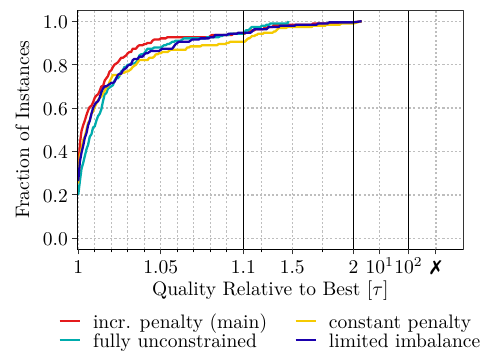}
	\else
		\tikzsetnextfilename{irregular_ufm_variants}%
		\input{tikz/irregular_ufm_variants}%
	\fi

	\end{minipage}

	\caption{Performance profiles comparing the solution quality for multiple variants of unconstrained local search on set I.
	This includes different combinations of unconstrained/constrained label propagation (LP) and FM (left)
    and alternative approaches to unconstrained FM (right).}
	\label{fig:design_quality}
\end{figure*}

The basic principle of unconstrained local search allows for a large number of possible designs.
Therefore, we discuss and evaluate several alternative approaches as well as important parameters.
Due to time constraints, we use five seeds for these experiments and restrict $k$ to powers of two.

\paragraph{Label Propagation and FM}
In the previous sections, we presented both an unconstrained variant of the FM algorithm and of label propagation refinement.
For previous size-constrained label propagation variants, it was observed that it provides no improved solution quality in addition to FM~\cite{MT-KAHYPAR-JOURNAL}.
However, this is different for the unconstrained case.
As shown in Figure~\ref{fig:design_quality} (left),
applying only the unconstrained FM algorithm results in significantly worse solution quality on irregular graphs.
A possible explanation is that the restriction caused by the penalty is a disadvantage on some instances,
while label propagation has no such restriction.
The results are similar if we use unconstrained label propagation in combination with constrained FM.
This indicates that the two unconstrained algorithms work best if used in combination.

\paragraph{Unconstrained FM Variants}
Further, there are multiple variants for an unconstrained FM algorithm that are arguably simpler than our approach,
which we compare in Figure~\ref{fig:design_quality} (right).
Recall that for all tested variants, we start with unconstrained FM but switch to constrained FM if the relative improvement is below a threshold of 0.002.
The \emph{fully unconstrained} FM variant allows any balance violating moves without penalties or other restrictions.
Perhaps surprisingly, the solution quality on set I is almost as good as our main configuration.
The downside of fully unconstrained FM is that it requires substantially more rebalancing.
The time spent for rebalancing during FM is roughly a factor 4 higher than with our main configuration in the geometric mean\footnote{\nolinebreak
We exclude timings close to zero to avoid deviations resulting from division by near-zero values. The overall result is stable for different thresholds (as well as without this step).},
resulting in an overall slowdown of 15\% for the FM component.
This suggests that balance-violating moves which would otherwise increase the overall cut size are compensated by performing more work in our high-quality rebalancing algorithm.

As explained in Section~\ref{s:integration}, in our main configuration, we \emph{increase} the penalty that is applied to balance violating moves over multiple rounds.
Instead, we could apply the same \emph{constant penalty} for all rounds (we use $\tau = 0.5$ which worked best in preliminary experiments).
The partitions computed by this variant are worse than our main configuration with a difference of 1.7\% in the geometric mean.
The third variant uses no penalty but instead allows only \emph{limited imbalance} per block.
More precisely, the localized FM search uses a maximum block weight of $\alpha_i \cdot L_{max}$ in round $i$.
We start with a large factor $\alpha_1 = 2$ and reduce the allowed imbalance in subsequent rounds,
using linear interpolation between $\alpha_1$ and $\alpha_{final} = 1.1$.
The quality of this variant is also similar to our main configuration,
but the FM component is 5\% slower.

Overall, the difference in solution quality between the variants is surprisingly small.
This indicates a high robustness of the basic principle of unconstrained FM, achieving good results in many different configurations.

\paragraph{Unconstrained FM Parameters}
An important parameter in the unconstrained FM algorithm is the threshold $t$,
which defines whether a node $v$ is available for rebalancing based on the fraction of incident edges that are connected to $\Partition(v)$ (see Section~\ref{s:penalties}).
A large value risks that useful nodes are excluded while a too small value might include nodes whose gain changes significantly during the round,
thereby making the penalty inaccurate.
Testing $t \in \{ 0.4, 0.7, 1 \}$ we found that $t = 0.7$ has slightly better solution quality on set I than the alternatives,
with a difference of 0.6\% and 0.2\% in the geometric mean compared to $0.4$ and $1$, respectively.
Further, based on preliminary experiments our algorithm switches to constrained FM if the relative improvement of a round is smaller than a threshold of 0.002.
In comparison to always using eight unconstrained and two constrained rounds, the evaluation shows that this has almost no impact on solution quality.
Meanwhile, it improves the running time of the FM component significantly,
by 27\% on set I and 15\% on set R.
Additionally, we tested a dynamic activation technique
that decides whether to use unconstrained or constrained FM based on the result of the first two rounds (see Section~\ref{s:integration}).
The results are mixed:
On set R, it improves the running time of the FM component by 9\% without decreasing the quality.
However, on set I the quality is 0.6\% worse in the geometric mean while the total running time is only improved by 0.4\%.
Due to these results and the added complexity of this technique we do not use it for the final configuration.

\begin{figure}[t]
	\ifpdfplots
		\includegraphics{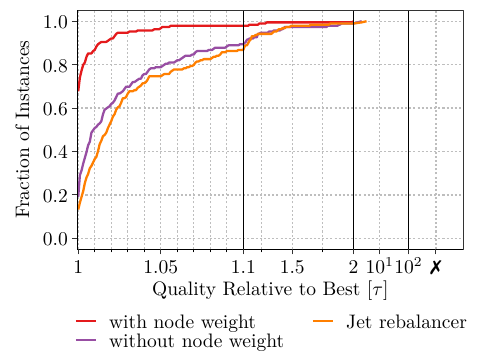}
	\else
		\tikzsetnextfilename{irregular_rebalancing}%
		\input{tikz/irregular_rebalancing}%
	\fi

	\caption{Performance profile comparing the solution quality for different rebalancing approaches on set I.}
	\label{fig:rebalancing}
	\vspace{8pt}
	\vfill
\end{figure}

\paragraph{Rebalancing and Move Sequence}
We compare different rebalancing algorithms in Figure~\ref{fig:rebalancing}.
First, our algorithm uses $\gain{\Partition}{m}/c(u)$ as priority for node $u$, where $m$ is the best move with balanced target block (see Section~\ref{s:rebalancing}).
However, since it is not clear that dividing by node weight is beneficial, we also tested a variant that directly uses the gain as priority.
In addition, we implemented the rebalancing algorithm that is used in Jet~\cite{JET}.
It is based on sorting the nodes into buckets according to their gain, afterwards processing the buckets in decreasing order of (negative) gain.
Since target blocks are selected in parallel without synchronizing the block weights,
moves to blocks within a \emph{deadzone} around the balance constraint are prohibited to avoid oscillation.
As shown, both alternative approaches achieve significantly worse quality on set I than our algorithm with included node weights.

After rebalancing, we construct an interleaved move sequence containing the FM moves and the rebalancing moves (see Section~\ref{s:move_sequence}).
In comparison to an approach where the rebalancing moves are just appended, this improves the quality by 0.8\% in the geometric mean.
The improvement is surprisingly small (it was larger in preliminary experiments), possibly because the rebalancer already achieves high quality on its own.

\subsection{Strong Scaling of the Rebalancing Algorithm}\label{subsec:scaling}

\begin{figure}
	\ifpdfplots
		\includegraphics{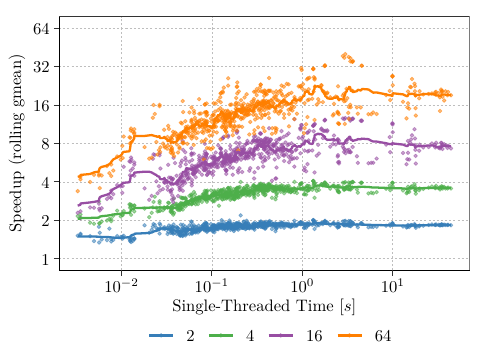}
	\else
		\tikzsetnextfilename{speedup_rebalancer}%
		\input{tikz/speedup_rebalancer}%
	\fi

	\caption{Strong scalability results for our rebalancing algorithm on set I. Each point corresponds to an instance (sorted by sequential time), and the line is a rolling window geometric mean.}\label{fig:strong-scaling}
\end{figure}

In \Cref{fig:strong-scaling}, we report instance-wise (and rolling window geometric mean) self-relative speedups for the rebalancing algorithm, using 2, 4, 16 and 64 cores.
With 2 and 4 cores we achieve near-linear speedups, whereas with 16 cores we see a drop-off to roughly 8x.
On graphs where sequential rebalancing takes more than 100ms, we achieve around 18x speedup on 64 cores in the geometric mean, and 22.93x on graphs that take longer than 1s.
These results are similar to the strong scalability results of the full framework~\cite{MT-KAHYPAR}.

\subsection{Solution Quality}\label{subsec:solution_quality}

\begin{figure*}[t]
	\begin{minipage}{0.49\textwidth}
	\ifpdfplots
		\includegraphics{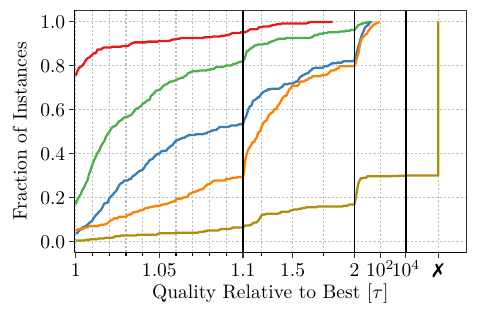}
	\else
		\tikzsetnextfilename{irregular_quality}%
		\input{tikz/irregular_quality}%
	\fi

	\end{minipage}
	\hfill
	\begin{minipage}{0.49\textwidth}
	\ifpdfplots
		\includegraphics{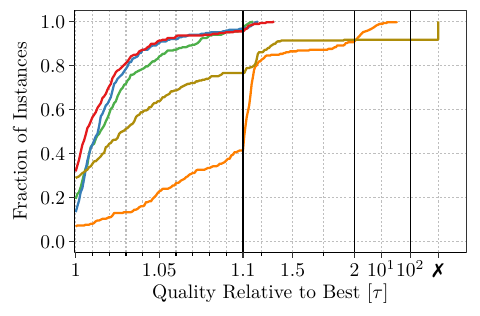}
	\else
		\tikzsetnextfilename{regular_quality}%
		\input{tikz/regular_quality}%
	\fi

	\end{minipage}

	\vfill
	\begin{minipage}{\textwidth}
	\ifpdfplots
		\includegraphics{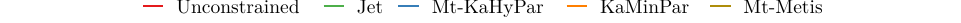}
	\else
		\tikzsetnextfilename{irregular_quality_legend}%
		\input{tikz/irregular_quality_legend}%
	\fi

	\end{minipage}
	\vfill
	\vspace{-2pt}

	\caption{Performance profiles comparing the solution quality of our unconstrained algorithm with different state-of-the-art partitioners on set I (left) and set R (right).}
	\label{fig:overall_quality}
\end{figure*}

\begin{figure*}[t]
	\hspace{18pt}
	\begin{minipage}{0.314\textwidth}
		\centering{\large{\textsc{Compression}}}
	\end{minipage}
	\hfill
	\begin{minipage}{0.314\textwidth}
		\centering{\large{\textsc{Web}}}
	\end{minipage}
	\hfill
	\begin{minipage}{0.314\textwidth}
		\centering{\large{\textsc{Brain}}}
	\end{minipage}
	\begin{minipage}{0.36\textwidth}
		\hspace{-2pt}
	\ifpdfplots
		\includegraphics{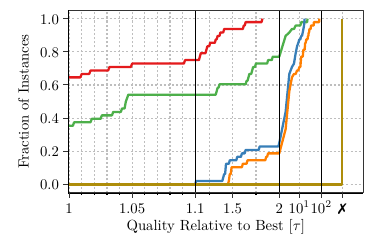}
	\else
		\tikzsetnextfilename{irregular_quality_recomp}%
		\input{tikz/irregular_quality_recomp}%
	\fi

	\end{minipage}
	\hfill
	\begin{minipage}{0.314\textwidth}
		\hspace{-3pt}
	\ifpdfplots
		\includegraphics{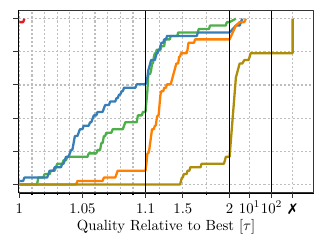}
	\else
		\tikzsetnextfilename{irregular_quality_web}%
		\input{tikz/irregular_quality_web}%
	\fi

	\end{minipage}
	\hfill
	\begin{minipage}{0.314\textwidth}
		\hspace{-6pt}
	\ifpdfplots
		\includegraphics{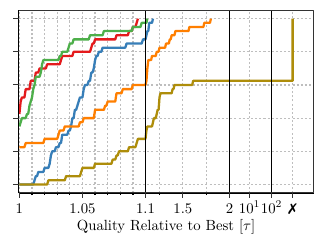}
	\else
		\tikzsetnextfilename{irregular_quality_brain}%
		\input{tikz/irregular_quality_brain}%
	\fi

	\end{minipage}

	\vspace{-8pt}
	\vfill
	\hspace{18pt}
	\begin{minipage}{0.314\textwidth}
		\centering{\large{\textsc{Social}}}
	\end{minipage}
	\hfill
	\begin{minipage}{0.314\textwidth}
		\centering{\large{\textsc{Wiki}}}
	\end{minipage}
	\hfill
	\begin{minipage}{0.314\textwidth}
		\centering{\large{\textsc{Artificial}}}
	\end{minipage}
	\begin{minipage}{0.36\textwidth}
		\hspace{-2pt}
	\ifpdfplots
		\includegraphics{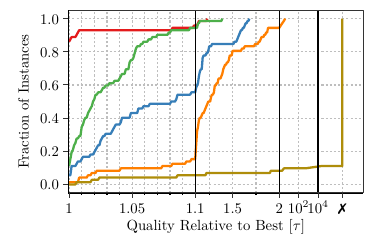}
	\else
		\tikzsetnextfilename{irregular_quality_social}%
		\input{tikz/irregular_quality_social}%
	\fi

	\end{minipage}
	\hfill
	\begin{minipage}{0.314\textwidth}
		\hspace{-3pt}
	\ifpdfplots
		\includegraphics{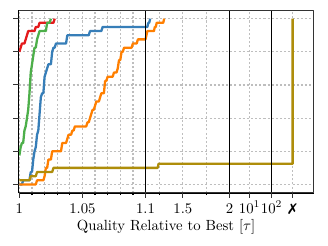}
	\else
		\tikzsetnextfilename{irregular_quality_wiki}%
		\input{tikz/irregular_quality_wiki}%
	\fi

	\end{minipage}
	\hfill
	\begin{minipage}{0.314\textwidth}
		\hspace{-6pt}
	\ifpdfplots
		\includegraphics{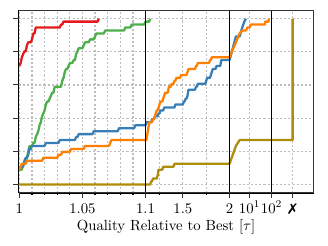}
	\else
		\tikzsetnextfilename{irregular_quality_random}%
		\input{tikz/irregular_quality_random}%
	\fi

	\end{minipage}

	\vspace{-8pt}
	\vfill
	\begin{minipage}{\textwidth}
	\ifpdfplots
		\includegraphics{experiments/irregular_quality_legend.pdf}
	\else
		\tikzsetnextfilename{irregular_quality_legend}%
		\input{tikz/irregular_quality_legend}%
	\fi

	\end{minipage}

	\caption{Performance profiles comparing the solution quality of our unconstrained algorithm with state-of-the-art partitioners for different classes of irregular graphs.}
	\label{fig:graph_classes}
\end{figure*}

In Figure~\ref{fig:overall_quality} we compare the solution quality of our unconstrained algorithm with other state-of-the-art parallel partitioners.
On 75\% of the irregular instances (set I), our algorithm finds the best solution of all considered partitioners.
Notably, we produce much better solutions than \mtkahypar{} although the only difference is that we replaced the label propagation and FM algorithms with unconstrained versions.
Among the competitors, \jet{} achieves the best quality on set I.
This is expected since \jet{} also allows temporary balance violations during the refinement.
However, in comparison to our algorithm, \jet{} has 9.6\% worse edge cut in the geometric mean.
Generally, on set I we observe a \emph{tail} in the result distribution where the partition found by the competing algorithms has a cut that is a factor of 2 or more larger than the best cut.
\mtmetis{} struggles to find balanced solutions. It produced only imbalanced solutions on 185 out of 304 instances (graph and $k$) in benchmark set I, and failed to produce any result on 28 instances\footnote{
	The execution either resulted in a segmentation fault or the 1TB of available main memory did not suffice for \mtmetis{}.
	This includes the large instances friendster, sk2005 and twitter2010 on set I and most of the kmer instances on set R.
}.

On regular graphs (set R), our algorithm also produces the best solutions of the compared partitioners,
although with a margin of less than 1\% in the geometric mean.
Both \mtkahypar{} and \jet{} achieve very similar quality to our algorithm,
while the remaining competitors produce significantly worse results.
Interestingly, \mtmetis{} performs much better on set R than on set I, finding at least one balanced solution on almost 90\% of instances.
It seems that the tendency of \mtmetis{} to produce imbalanced results is amplified by irregular graphs.

Additionally, we evaluate the influence of the different graph classes contained in set I.
The results are depicted in Figure~\ref{fig:graph_classes}.
Our unconstrained algorithm has superior quality for each graph class,
except brain graphs where the result quality is roughly equivalent to Jet.
We find the largest differences compared to the second best algorithm on web graphs and on compression graphs.
Our algorithm computes for 98\% of the web graph instances the best solution, with an overall improvement of 17\% compared to the best competitor (Jet).
For compression graphs, the difference is even 43\% in the geometric mean, since the results of all competitors exhibit a large tail with substantially worse quality.
Another interesting class are the artificial graphs,
since Jet produces comparatively good results (2.6\% worse than our algorithm) but all remaining competitors have substantially inferior quality, by at least 60\% in the geometric mean.

\subsection{Running Time}\label{subsec:time}

\begin{table}
	\vfill
	\vspace{-3pt}
	\caption{Geometric mean running time of the partitioners in seconds ($^*$excluding failed runs for Mt-Metis).}
	\label{tab:running_times}

	\vspace{6pt}
	\begin{tabular}{lrr}
		Algorithm & Time on set I & Time on set R\\
		\midrule
		Unconstrained & 9.77 & 2.74\\
		\midrule
		\jet{} & 9.07 & 2.99\\
		\mtkahypar{} & 8.67 & 2.54\\
		\kaminpar{} & 2.61 & 1.55\\
		\mtmetis{}$^*$ & 7.85 & 1.73
	\end{tabular}
	\vfill
\end{table}

The running times of our algorithm and the competitors are summarized in Table~\ref{tab:running_times}.
Our unconstrained algorithm has only small overhead in comparison to \mtkahypar{},
it is 12.7\% slower on set I and 8\% slower on set R.
Compared to \jet{}, it is only 7.7\% slower on set I and even 8.8\% faster on set R.
\mtmetis{} is also faster than our algorithm, but no more than a factor of 1.6.
\kaminpar{} is much faster on both instance sets but this comes at the cost of solution quality.
We also investigated which components of our algorithm cause the observed running time overhead.
As expected, we can attribute most overhead to the rebalancing that is necessary for unconstrained refinement.
If we subtract the rebalancing time for each run on set I,
the resulting geometric mean is only 1.3\% larger than the running time of \mtkahypar{}.
On set R, the same technique attributes roughly a quarter of the overhead to the rebalancing.

These results demonstrate that unconstrained local search has only small running time overhead in comparison to constrained techniques.
Our solution quality exceeds all competitors by a large margin, which clearly results in a worthwhile trade-off.

\section{Conclusion}\label{s:conclusion}

We develop new refinement techniques for graph partitioning which combine \emph{unconstrained moves} with a separate \emph{rebalancing} step later on.
Among multiple tested configurations,
the best results are achieved by combining unconstrained label propagation with an unconstrained FM variant that uses approximate penalties for balance-violating moves,
as well as a high-quality rebalancing algorithm.
Our experiments demonstrate that unconstrained refinement finds substantially better solutions than current state-of-the-art partitioners,
due to its capability to escape local optima.
The effect is especially pronounced on irregular graphs with skewed degree distribution,
where clusters of high degree nodes have large influence on the cut size.

The notion of unconstrained local search opens a vast space of new design options that are not available to size-constrained approaches.
Of key interest are improved heuristics for determining when to rebalance, how to penalize balance-violating moves and when to prefer constrained refinement.
While we tried several approaches in this paper already, there are more choices to explore.
In addition, we plan to investigate whether improvements in coarsening and initial partitioning can amplify the effect of unconstrained refinement.
Finally, we would like to evaluate unconstrained refinement for the more general hypergraph partitioning problem.

\vspace{.2cm}
\paragraph{Acknowledgements} We thank Mike Gilbert for fruitful discussions and his review of our Jet refinement implementation.
This project has received funding from the European Research Council (ERC) under the European Union’s Horizon 2020 research and innovation programme (grant agreement No. 882500).\\

\centering{\includegraphics[width=0.5\linewidth]{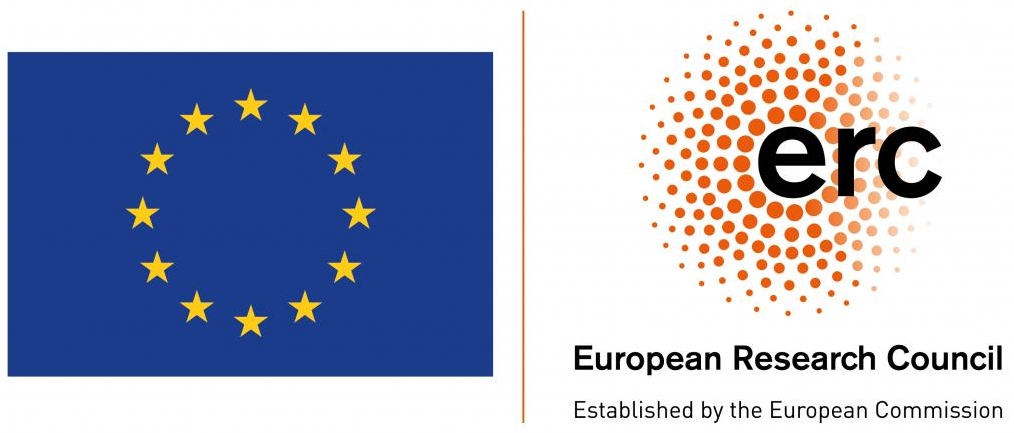}}

\newpage
\bibliography{mt_kahypar}

\end{document}